\documentclass[ aip, jmp, preprint,]{revtex4-1}
\usepackage{amsmath}
\usepackage{graphicx}
\usepackage{dcolumn}
\usepackage{bm}

\begin{document}

\preprint{}
\title[]{Identity for the DFT correlation functional.}
\author{Daniel P. Joubert}
\email{daniel.joubert2@wits.ac.za}
\affiliation{Centre for Theoretical Physics, University of the Witwatersrand, PO Wits
2050, Johannesburg, South Africa}
\date{\today }

\begin{abstract}
It is shown that the electron density functional correlation functional
satisfies%
\begin{eqnarray}
&&E_{c}^{\gamma }\left[ \rho _{N}\right] -E_{c}^{\gamma }\left[ \rho
_{N-1}^{\gamma }\right]   \notag \\
&&+\gamma \int_{0}^{\gamma }d\lambda \int d^{3}r\left( v_{c}\left( \left[
\rho _{N-1}^{\lambda }\right] ;\mathbf{r}\right) -v_{c}\left( \left[ \rho
_{N}\right] ;\mathbf{r}\right) \right) \frac{\partial }{\partial \lambda }%
\rho _{N-1}^{\lambda }\left( \mathbf{r}\right)   \notag \\
&=&\int d^{3}r^{\prime }\left( \rho _{N}\left( \mathbf{r}^{\prime }\right)
-\rho _{N-1}^{\gamma }\left( \mathbf{r}^{\prime }\right) \right) v_{c}\left( %
\left[ \rho _{N}\right] ;\mathbf{r}\right) \notag .
\end{eqnarray}%
$\rho _{N}\left( \mathbf{r}\right) $ and $\rho _{N-1}^{\gamma }\left(
\mathbf{r}\right) $ are $N$-electron and $(N-1)$-electron densities
determined from the same adiabatically scaled Hamiltonian of the interacting
electron system with $\gamma $ the scaling parameter of the
electron-electron interaction strength.
\end{abstract}


\maketitle














\section{ Introduction}

The Kohn-Sham (KS) formulation \cite{KohnSham:65} of Density Functional
Theory (DFT) \cite{HohenbergKohn:64,ParrYang:bk89,DreizlerGross:bk90} is one of the most important tools for
the calculation of electronic structure of molecules and solids. In all
practical applications of DFT, however, approximations to the exact
functionals have to be made \cite%
{AdriennRuzsinszky2011,JohnP.Perdew2009,GaborI.Csonka2009a,Staroverov2004,TaoPerdew03}.
Exact relations for density functionals and density functional derivatives
can play an important role in the development of accurate approximations to
the exact functionals\cite{JohnP.Perdew2005}. In this paper the following equation satisfied by the
exact correlation energy and potential is derived:%
\begin{eqnarray}
&&E_{c}^{\gamma }\left[ \rho _{N}\right] -E_{c}^{\gamma }\left[ \rho
_{N-1}^{\gamma }\right]   \notag \\
&&+\gamma \int_{0}^{\gamma }d\lambda \int d^{3}r\left( v_{c}\left( \left[
\rho _{N-1}^{\lambda }\right] ;\mathbf{r}\right) -v_{c}\left( \left[ \rho
_{N}\right] ;\mathbf{r}\right) \right) \frac{\partial }{\partial \lambda }%
\rho _{N-1}^{\lambda }\left( \mathbf{r}\right)   \notag \\
&=&\int d^{3}r^{\prime }\left( \rho _{N}\left( \mathbf{r}^{\prime }\right)
-\rho _{N-1}^{\gamma }\left( \mathbf{r}^{\prime }\right) \right) v_{c}\left( %
\left[ \rho _{N}\right] ;\mathbf{r}\right) .  \label{ID1}
\end{eqnarray}

Here $\rho _{N}$ and $\rho _{N-1}^{\gamma }$ are the ground state charge
densities of an interacting $N$ and $\left( N-1\right) $ electron system of
the same Hamiltonian with multiplicative external potential $v_{\text{ext}%
}^{\gamma }\left( [\rho _{N}]\right) $. The potential $v_{\text{ext}%
}^{\gamma }\left( [\rho _{N}]\right) $ is constructed to keep the charge
density of the $N$ electron system independent of the coupling strength
parameter $\gamma $ \cite%
{HarrisJones:74,LangrethPerdew:75,LangrethPerdew:77,GunnarsonLundqvist:76}
that scales the electron-electron interaction strength. At $\gamma =1$ full
strength Coulomb interaction between electrons is included and the external
potential $v_{\text{ext}}^{\gamma }\left( [\rho _{N}]\right) $ is the
external potential of the fully interacting system, while $\gamma =0$
corresponds to the non-interacting Kohn-Sham potential.

It is well known that the correlation energy $E_{c}^{\gamma }\left[ \rho _{N}%
\right] $ at coupling strength $\gamma $ can be expressed as the integral
\cite%
{HarrisJones:74,LangrethPerdew:75,LangrethPerdew:77,GunnarsonLundqvist:76}
\begin{equation}
E_{c}^{\gamma }\left[ \rho _{N}\right] =\int_{0}^{\gamma }d\lambda
V_{ee}^{\lambda }\left[ \rho _{N}\right] -\gamma E_{hx}\left[ \rho _{N}%
\right]
\end{equation}%
where $V_{ee}^{\gamma }\left[ \rho _{N}\right] $ is the electron-electron Coulomb
interaction energy and $E_{hx}\left[ \rho _{N}%
\right] $ is the sum of the Hartree and exchange energy. An
alternative way of expressing this relationship is%
\begin{equation}
E_{c}^{\gamma }\left[ \rho _{N}\right] =\int_{0}^{\gamma }d\lambda \frac{1}{%
\lambda }\left( E_{c}^{\lambda }\left[ \rho _{N}\right] -T_{c}^{\lambda }%
\left[ \rho _{N}\right] \right) .
\end{equation}%
This equation shows that the there is a close relationship between the
correlation energy $E_{c}^{\gamma }\left[ \rho _{N}\right] $ at $\gamma $
and the correlation energy and the correlation part of the kinetic energy 
$T_{c}^{\gamma }\left[ \rho _{N}\right] $ 
for all coupling strengths between $0$ and $\gamma $. This expression is not solely 
in terms of the correlation energy.  Equation (\ref{ID1}), on the other hand,
provides a relation entirely in terms of correlation functionals. It is remarkable 
in that it not only relates the correlation functionals at 
different densities of the same Hamiltonian, but indicates a relationship between the 
correlation functionals at all coupling strengths between $0$ and $\gamma .$
Equation (\ref{ID1}) can serve as a test for approximate correlation
functionals provided accurate densities are used.

\section{Proof}

In the adiabatic connection approach \cite%
{HarrisJones:74,LangrethPerdew:75,LangrethPerdew:77,GunnarsonLundqvist:76}
of the constrained minimization formulation of density functional theory
\cite{HohenbergKohn:64,KohnSham:65,Levy:79,LevyPerdew:85} the Hamiltonian $%
\hat{H}^{\gamma }$ for a system of $N$ electrons is given by
\begin{equation}
\hat{H}^{\gamma }=\hat{T}+\gamma \hat{V}_{ee}+\hat{v}_{\text{ext}}^{\gamma
}\left[ \rho _{N}\right] .  \label{a3}
\end{equation}%
Atomic units, $\hbar =e=m=1$ are used throughout. $\hat{T}$ is the kinetic
energy operator,%
\begin{equation}
\hat{T}=-\frac{1}{2}\sum_{i=1}^{N}\nabla _{i}^{2},  \label{a4}
\end{equation}%
\ and $\gamma \hat{V}_{ee\text{ }}$is the scaled electron-electron interaction operator%
\begin{equation}
\gamma \hat{V}_{ee}=\gamma \sum_{i<j}^{N}\frac{1}{\left\vert \mathbf{r}_{i}-%
\mathbf{r}_{j}\right\vert }.  \label{a2}
\end{equation}%
The external potential
\begin{equation}
\hat{v}_{\text{ext}}^{\gamma }\left[ \rho _{N}\right] =\sum_{i=1}^{N}v_{%
\text{ext}}^{\gamma }\left( \left[ \rho _{N}\right] ;\mathbf{r}_{i}\right) ,
\label{a1}
\end{equation}%
is constructed to keep the charge density of the $N$-electron system fixed at $\rho _{N}\left( \mathbf{r%
}\right) ,$ the ground state charge density of the fully interacting system (%
$\gamma =1$), for all values of the coupling constant $\gamma .$ The
external potential has the form \cite{LevyPerdew:85,GorlingLevy:93}
\begin{align}
v_{\text{ext}}^{\gamma }(\left[ \rho _{N}\right] ;\mathbf{r})& =\left(
1-\gamma \right) v_{hx}([\rho _{N}];\mathbf{r})  \notag \\
& +v_{c}^{1}([\rho _{N}];\mathbf{r)}-v_{c}^{\gamma }([\rho _{N}];\mathbf{r)+}%
v_{\text{ext}}^{1}(\left[ \rho _{N}\right] ;\mathbf{r}),  \label{e1}
\end{align}%
where $v_{\text{ext}}^{1}(\left[ \rho _{N}\right] ;\mathbf{r})=v_{\text{ext}%
}\left( \mathbf{r}\right) $ is the external potential at full coupling
strength, $\gamma =1,$ and $v_{\text{ext}}^{0}(\left[ \rho _{N}\right] ;%
\mathbf{r})$ is the non-interacting Kohn-Sham potential. The exchange plus
Hartree potential \cite{ParrYang:bk89,DreizlerGross:bk90} $v_{hx}([\rho _{N}];%
\mathbf{r}),$ is independent of $\gamma ,$ while the correlation potential $%
v_{c}^{\gamma }([\rho _{N}];\mathbf{r)}$ depends in the scaling parameter $%
\gamma .$

The energy functional $F^{\gamma }[\rho ]$ is defined as \cite%
{Levy:79,ParrYang:bk89,DreizlerGross:bk90} the sum of the kinetic and interaction energy,
\begin{eqnarray}
F^{\gamma }[\rho ] &=&\left\langle \Psi _{\rho }^{\gamma }\left\vert \hat{T}%
+\gamma \hat{V}_{ee}\right\vert \Psi _{\rho }^{\gamma }\right\rangle   \notag
\\
&=&\min_{\Psi \rightarrow \rho }\left\langle \Psi \left\vert \hat{T}+\gamma
\hat{V}_{ee}\right\vert \Psi \right\rangle ,  \label{b1}
\end{eqnarray}%
where according to the Levy constrained minimization definition\cite{Levy:79}%
, the wavefunction $\left\vert \Psi _{\rho }^{\gamma }\right\rangle $ yields
the density $\rho $ and minimizes $\left\langle \Psi \left\vert \hat{T}%
+\gamma \hat{V}_{ee}\right\vert \Psi \right\rangle .$ For $w$-representable
densities\cite{Levy:79,ParrYang:bk89,DreizlerGross:bk90}, the densities are
derived from the groundstate eigenfunctions of a Hamiltonian:%
\begin{eqnarray}
\hat{H}^{\gamma }\left\vert \Psi _{\rho ^{\gamma }}^{\gamma }\right\rangle
&=&E_{M}^{\gamma }\left\vert \Psi _{\rho ^{\gamma }}^{\gamma }\right\rangle
\notag \\
\hat{H}^{\gamma } &=&\hat{T}+\gamma \hat{V}_{ee}+\hat{v}_{\text{ext}%
}^{\gamma }\left[ \rho \right] .  \label{h1}
\end{eqnarray}%
Note that by construction of $v_{\text{ext}}^{\gamma }\left( \left[ \rho _{N}%
\right] ;\mathbf{r}\right) ,$ Eq. (\ref{e1}) $\rho _{N}=\rho _{N}^{1}$, is
independent of $\gamma ,$ but the groundstate density of the $\left(
N-1\right) $-electron system $\rho _{N-1}^{\gamma }$, is a function of $%
\gamma.$    $F^{\gamma }[\rho ]$ can be decomposed as \cite%
{ParrYang:bk89,DreizlerGross:bk90}%
\begin{equation}
F^{\gamma }[\rho ]=T^{0}\left[ \rho \right] +\gamma E_{hx}\left[ \rho \right]
+E_{c}^{\gamma }\left[ \rho \right] ,  \label{b2}
\end{equation}%
The correlation energy $E_{c}^{\gamma }\left[ \rho \right] $ is defined as%
\cite{LevyPerdew:85}
\begin{eqnarray}
E_{c}^{\gamma }\left[ \rho \right]  &=&\left\langle \Psi _{\rho }^{\gamma
}\left\vert \hat{T}+\gamma \hat{V}_{ee}\right\vert \Psi _{\rho }^{\gamma
}\right\rangle   \notag \\
&&-\left\langle \Psi _{\rho }^{0}\left\vert \hat{T}+\gamma \hat{V}%
_{ee}\right\vert \Psi _{\rho }^{0}\right\rangle ,  \label{ec1}
\end{eqnarray}%
where $\left\vert \Psi _{\rho }^{0}\right\rangle $ is the Kohn-Sham
independent particle groundstate wavefunction that yields the same density
as the interacting system at coupling strength $\gamma .$ $E_{hx}\left[ \rho %
\right] $ is the sum of the Hartree and exchange energy%
\begin{equation}
E_{hx}\left[ \rho \right] =\left\langle \Psi _{\rho }^{0}\left\vert \hat{V}%
_{ee}\right\vert \Psi _{\rho }^{0}\right\rangle,   \label{b3}
\end{equation}%
and the non-interacting kinetic energy functional $T^{0}\left[ \rho \right] $ is given by%
\begin{equation}
T^{0}\left[ \rho \right] =\left\langle \Psi _{\rho }^{0}\left\vert \hat{T}%
\right\vert \Psi _{\rho }^{0}\right\rangle .  \label{b4}
\end{equation}%
The full kinetic energy
\begin{eqnarray}
T^{\gamma }\left[ \rho \right]  &=&\left\langle \Psi _{\rho }^{\gamma
}\left\vert \hat{T}\right\vert \Psi _{\rho }^{\gamma }\right\rangle   \notag
\\
&=&T^{0}\left[ \rho \right] +T_{c}^{\gamma }\left[ \rho \right] ,  \label{b5}
\end{eqnarray}%
with the correlation part of the kinetic%
\begin{equation}
T_{c}^{\gamma }\left[ \rho \right] =\left\langle \Psi _{\rho }^{\gamma
}\left\vert \hat{T}\right\vert \Psi _{\rho }^{\gamma }\right\rangle
-\left\langle \Psi _{\rho }^{0}\left\vert \hat{T}\right\vert \Psi _{\rho
}^{0}\right\rangle .  \label{ec2}
\end{equation}%
Assuming that $F^{\gamma }[\rho ]$ is defined for non-integer electrons \cite%
{PPLB:82,ParrYang:bk89,DreizlerGross:bk90}, at the solution point%
\begin{equation}
\frac{\delta F^{\gamma }\left[ \rho \right] }{\delta \rho \left( \mathbf{r}%
\right) }+v_{\text{ext}}^{\gamma }\left( \left[ \rho \right] ;\mathbf{r}%
\right) =\mu   \label{b6}
\end{equation}%
where $\mu $ is the chemical potential.

In a recent paper \cite{Joubert2011b} the author proved that the correlation
part of the kinetic energy satisfies%
\begin{eqnarray}
&&T_{c}^{\gamma }\left[ \rho _{N}\right] -T_{c}^{\gamma }\left[ \rho
_{N-1}^{\gamma }\right]  \notag \\
&=&\int d^{3}r^{\prime }\left( \rho _{N}\left( \mathbf{r}^{\prime }\right)
-\rho _{N-1}^{\gamma }\left( \mathbf{r}^{\prime }\right) \right) \frac{%
\delta T_{c}^{\gamma }\left[ \rho _{N}\right] }{\delta \rho _{N}\left(
\mathbf{r}\right) }.  \label{id4}
\end{eqnarray}%
In appendix A, Eq. (\ref{apf}), it is shown that%

\begin{eqnarray}
&&T_{c}^{\gamma }\left[ \rho _{N-1}^{\gamma }\right]  \notag \\
&=&E_{c}^{\gamma }\left[ \rho _{N-1}^{\gamma }\right] -\gamma \frac{\partial
}{\partial \gamma }E_{c}^{\gamma }\left[ \rho _{N-1}^{\gamma }\right]  \notag
\\
&&+\gamma \int d^{3}r\frac{\partial \rho _{N-1}^{\gamma }\left( \mathbf{r}%
\right) }{\partial \gamma }\left( v_{c}^{\gamma }\left( \left[ \rho _{N}%
\right] ;\mathbf{r}\right) +\gamma v_{hx}\left( \left[ \rho _{N}\right] ;%
\mathbf{r}\right) -\gamma v_{hx}\left( \left[ \rho _{N-1}^{\gamma }\right] ;%
\mathbf{r}\right) \right) ,  \label{p1}
\end{eqnarray}%
where $v_{c}^{\gamma }\left( \left[ \rho \right] ;\mathbf{r}\right) =\frac{%
\delta }{\delta \rho \left( \mathbf{r}\right) }E_{c}^{\gamma }\left[ \rho %
\right] $ and $v_{hx}\left( \left[ \rho _{N}\right] ;\mathbf{r}\right) =%
\frac{\delta }{\delta \rho \left( \mathbf{r}\right) }E_{hx}\left[ \rho %
\right] $. It follows that%
\begin{eqnarray}
&&\frac{\partial }{\partial \gamma }T_{c}^{\gamma }\left[ \rho
_{N-1}^{\gamma }\right]  \notag \\
&=&-\gamma \frac{\partial ^{2}}{\partial \gamma ^{2}}E_{c}^{\gamma }\left[
\rho _{N-1}^{\gamma }\right]  \notag \\
&&+\int d^{3}r\frac{\partial \rho _{N-1}^{\gamma }\left( \mathbf{r}\right) }{%
\partial \gamma }\left( v_{c}^{\gamma }\left( \left[ \rho _{N}\right] ;%
\mathbf{r}\right) +\gamma v_{hx}\left( \left[ \rho _{N}\right] ;\mathbf{r}%
\right) -\gamma v_{hx}\left( \left[ \rho _{N-1}^{\gamma }\right] ;\mathbf{r}%
\right) \right)  \notag \\
&&+\gamma \frac{\partial }{\partial \gamma }\int d^{3}r\frac{\partial \rho
_{N-1}^{\gamma }\left( \mathbf{r}\right) }{\partial \gamma }\left(
v_{c}^{\gamma }\left( \left[ \rho _{N}\right] ;\mathbf{r}\right) +\gamma
v_{hx}\left( \left[ \rho _{N}\right] ;\mathbf{r}\right) -\gamma v_{hx}\left( %
\left[ \rho _{N-1}^{\gamma }\right] ;\mathbf{r}\right) \right) .  \label{p2}
\end{eqnarray}%
A similar expression can be derived for $\frac{\partial }{\partial \gamma }%
T_{c}^{\gamma }\left[ \rho _{N}\right] $ but since $\rho _{N}$ is
independent of $\gamma ,$ the result is simply%
\begin{equation}
\frac{\partial }{\partial \gamma }T_{c}^{\gamma }\left[ \rho _{N}\right]
=-\gamma \frac{\partial ^{2}}{\partial \gamma ^{2}}E_{c}^{\gamma }\left[
\rho _{N-1}^{\gamma }\right] .  \label{p3}
\end{equation}%
Taking the derivative of (\ref{p1}) with respect to $\gamma $, using (\ref%
{p2}) and (\ref{p3}) and some algebra, it can be shown that

\begin{eqnarray}
&&\frac{\partial ^{2}}{\partial \gamma ^{2}}E_{c}^{\gamma }\left[ \rho _{N}%
\right] -\frac{\partial ^{2}}{\partial \gamma ^{2}}E_{c}^{\gamma }\left[
\rho _{N-1}^{\gamma }\right]   \notag \\
&&+\int d^{3}r\frac{\partial }{\partial \gamma }\rho _{N-1}^{\gamma }\left(
\mathbf{r}\right) \left( v_{ux}\left( \left[ \rho _{N}\right] ;\mathbf{r}%
\right) -v_{ux}\left( \left[ \rho _{N-1}^{\gamma }\right] ;\mathbf{r}\right)
\right)   \notag \\
&&+\frac{\partial }{\partial \gamma }\int d^{3}r\frac{\partial \rho
_{N-1}^{\gamma }\left( \mathbf{r}\right) }{\partial \gamma }\left(
v_{c}^{\gamma }\left( \left[ \rho _{N}\right] ;\mathbf{r}\right) +\gamma
v_{ux}\left( \left[ \rho _{N}\right] ;\mathbf{r}\right) -\gamma v_{ux}\left( %
\left[ \rho _{N-1}^{\gamma }\right] ;\mathbf{r}\right) \right)   \notag \\
&=&\frac{\partial }{\partial \gamma }\int d^{3}r^{\prime }\left( \rho
_{N}\left( \mathbf{r}^{\prime }\right) -\rho _{N-1}^{\gamma }\left( \mathbf{r%
}^{\prime }\right) \right) \left( \frac{\delta }{\delta \rho _{N}\left(
\mathbf{r}\right) }\frac{\partial }{\partial \gamma }E_{c}^{\gamma }\left[
\rho _{N}\right] \right)   \label{p4}
\end{eqnarray}%
This equation can be integrated with respect to $\gamma $ and since by
definition $E_{c}^{0}\left[ \rho \right] =\left. \frac{\partial }{\partial
\gamma }E_{c}^{\gamma }\left[ \rho \right] \right\vert _{\gamma =0}=0,$%
\begin{eqnarray}
&&\frac{\partial }{\partial \gamma }E_{c}^{\gamma }\left[ \rho _{N}\right] -%
\frac{\partial }{\partial \gamma }E_{c}^{\gamma }\left[ \rho _{N-1}^{\gamma }%
\right]   \notag \\
&&+\int_{0}^{\gamma }d\gamma \int d^{3}r\frac{\partial }{\partial \gamma }%
\rho _{N-1}^{\gamma }\left( \mathbf{r}\right) \left( v_{ux}\left( \left[
\rho _{N}\right] ;\mathbf{r}\right) -v_{ux}\left( \left[ \rho _{N-1}^{\gamma
}\right] ;\mathbf{r}\right) \right)   \notag \\
&&+\int d^{3}r\frac{\partial \rho _{N-1}^{\gamma }\left( \mathbf{r}\right) }{%
\partial \gamma }\left( v_{c}^{\gamma }\left( \left[ \rho _{N}\right] ;%
\mathbf{r}\right) +\gamma v_{ux}\left( \left[ \rho _{N}\right] ;\mathbf{r}%
\right) -\gamma v_{ux}\left( \left[ \rho _{N-1}^{\gamma }\right] ;\mathbf{r}%
\right) \right)   \notag \\
&=&\int d^{3}r^{\prime }\left( \rho _{N}\left( \mathbf{r}^{\prime }\right)
-\rho _{N-1}^{\gamma }\left( \mathbf{r}^{\prime }\right) \right) \left(
\frac{\delta }{\delta \rho _{N}\left( \mathbf{r}\right) }\frac{\partial }{%
\partial \gamma }E_{c}^{\gamma }\left[ \rho _{N}\right] \right) .  \label{p5}
\end{eqnarray}

From Eq. (\ref{apf}) applied the ${\gamma }$ invariant $N$-electron density,  it follows that

\begin{equation}
\frac{\partial }{\partial \gamma }E_{c}^{\gamma }\left[ \rho _{N}\right] =%
\frac{E_{c}^{\gamma }\left[ \rho _{N}\right] -T_{c}^{\gamma }\left[ \rho _{N}%
\right] }{\gamma }.  \label{p7}
\end{equation}%
Substitute (\ref{apf}) and (\ref{p7}) in (\ref{p5}) with the result%
\begin{eqnarray}
&&E_{c}^{\gamma }\left[ \rho _{N}\right] -E_{c}^{\gamma }\left[ \rho
_{N-1}^{\gamma }\right]   \notag \\
&&+\gamma \int d\gamma \int d^{3}r\frac{\partial }{\partial \gamma }\rho
_{N-1}^{\gamma }\left( \mathbf{r}\right) \left( v_{ux}\left( \left[ \rho _{N}%
\right] ;\mathbf{r}\right) -v_{ux}\left( \left[ \rho _{N-1}^{\gamma }\right]
;\mathbf{r}\right) \right)   \notag \\
&=&\int d^{3}r^{\prime }\left( \rho _{N}\left( \mathbf{r}^{\prime }\right)
-\rho _{N-1}^{\gamma }\left( \mathbf{r}^{\prime }\right) \right) \frac{%
\delta }{\delta \rho _{N}\left( \mathbf{r}\right) }E_{c}^{\gamma }\left[
\rho _{N}\right] .  \label{p8}
\end{eqnarray}%
This equation still contains reference to the Hartree and exchange
potentials. In order to find an equation that explicitly only contains
correlation energy functionals and potentials consider the following. In a
recent paper\cite{Joubert2011b} the author also showed that the full kinetic energy
functional satisfies
\begin{eqnarray}
&&T^{\gamma }\left[ \rho _{N}\right] -T^{\gamma }\left[ \rho _{N-1}^{\gamma }%
\right]   \notag \\
&=&\int d^{3}r^{\prime }\left( \rho _{N}\left( \mathbf{r}^{\prime }\right)
-\rho _{N-1}^{\gamma }\left( \mathbf{r}^{\prime }\right) \right) \frac{%
\delta T^{\gamma }\left[ \rho _{N}\right] }{\delta \rho _{N}\left( \mathbf{r}%
\right) }.  \label{p9}
\end{eqnarray}%
Taking into account that
\begin{equation}
T^{\gamma }\left[ \rho \right] =T^{0}\left[ \rho \right] +T_{c}^{\gamma }%
\left[ \rho \right]   \label{p10}
\end{equation}%
and that $\rho _{N-1}^{\gamma }$ is a function of $\gamma $\cite%
{ParrYang:bk89}, the $\gamma $ derivative of (\ref{p9}) can be written as

\begin{eqnarray}
&&\frac{\partial }{\partial \gamma }\left( T_{c}^{\gamma }\left[ \rho _{N}%
\right] -T_{c}^{\gamma }\left[ \rho _{N-1}^{\gamma }\right] \right)  \notag
\\
&=&\frac{\partial }{\partial \gamma }\left( \int d^{3}r^{\prime }\left( \rho
_{N}\left( \mathbf{r}^{\prime }\right) -\rho _{N-1}^{\gamma }\left( \mathbf{r%
}^{\prime }\right) \right) \frac{\delta T_{c}^{\gamma }\left[ \rho _{N}%
\right] }{\delta \rho _{N}\left( \mathbf{r}\right) }\right)  \notag \\
&&+\int d^{3}r\frac{\partial \rho _{N-1}^{\gamma }\left( \mathbf{r}\right)
}{\partial \gamma }\left( \frac{\delta T^{0}\left[ \rho _{N-1}^{\gamma }%
\right] }{\delta \rho _{N-1}^{\gamma }\left( \mathbf{r}\right) }-\frac{%
\delta T^{0}\left[ \rho _{N}\right] }{\delta \rho _{N}\left( \mathbf{r}%
\right) }\right) .  \label{p11}
\end{eqnarray}%
Comparison with (\ref{id4}) shows that%
\begin{equation}
\int d^{3}r\frac{\partial \rho _{N-1}^{\gamma }\left( \mathbf{r}\right) }{%
\partial \gamma }\left( \frac{\delta T^{0}\left[ \rho _{N-1}^{\gamma }\right]
}{\delta \rho _{N-1}^{\gamma }\left( \mathbf{r}\right) }-\frac{\delta T^{0}%
\left[ \rho _{N}\right] }{\delta \rho _{N}\left( \mathbf{r}\right) }\right)
=0.  \label{p12}
\end{equation}%
By construction, from (\ref{b6}) and (\ref{b2})%
\begin{eqnarray}
\frac{\delta T^{0}\left[ \rho _{N}\right] }{\delta \rho _{N}\left( \mathbf{r}%
\right) }+v_{c}^{\gamma }\left( \left[ \rho _{N}\right] ;\mathbf{r}\right)
+\gamma v_{ux}\left( \left[ \rho _{N}\right] ;\mathbf{r}\right) +v_{\text{ext%
}}^{\gamma }\left( \left[ \rho _{N}\right] ;\mathbf{r}\right) &=&\mu
_{N}^{\gamma }  \notag \\
\frac{\delta T^{0}\left[ \rho _{N}\right] }{\delta \rho _{N}\left( \mathbf{r}%
\right) }+v_{c}^{\gamma }\left( \left[ \rho _{N-1}^{\gamma }\right] ;\mathbf{%
r}\right) +\gamma v_{ux}\left( \left[ \rho _{N-1}^{\gamma }\right] ;\mathbf{r%
}\right) +v_{\text{ext}}^{\gamma }\left( \left[ \rho _{N-1}^{\gamma }\right]
;\mathbf{r}\right) &=&\mu _{N-1}^{\gamma }  \label{p13}
\end{eqnarray}%
and hence%
\begin{eqnarray}
&&\frac{\delta T^{0}\left[ \rho _{N-1}^{\gamma }\right] }{\delta \rho
_{N-1}^{\gamma }\left( \mathbf{r}\right) }-\frac{\delta T^{0}\left[ \rho _{N}%
\right] }{\delta \rho _{N}\left( \mathbf{r}\right) }  \notag \\
&=&v_{c}^{\gamma }\left( \left[ \rho _{N-1}^{\gamma }\right] ;\mathbf{r}%
\right) +\gamma v_{ux}\left( \left[ \rho _{N-1}^{\gamma }\right] ;\mathbf{r}%
\right) -v_{c}^{\gamma }\left( \left[ \rho _{N}\right] ;\mathbf{r}\right)
-\gamma v_{ux}\left( \left[ \rho _{N}\right] ;\mathbf{r}\right) +\mu
_{N}^{\gamma }-\mu _{N-1}^{\gamma }.  \label{p14}
\end{eqnarray}%
Now%
\begin{equation}
\frac{\partial }{\partial \gamma }\left( N-1\right) =\int d^{3}r\frac{%
\partial \rho _{N-1}^{\gamma }\left( \mathbf{r}\right) }{\partial \gamma }=0.
\label{p15}
\end{equation}%
From (\ref{p15}), (\ref{p14}), (\ref{p12}) and (\ref{p8}) the main result of this paper follows
\begin{eqnarray}
&&E_{c}^{\gamma }\left[ \rho _{N}\right] -E_{c}^{\gamma }\left[ \rho
_{N-1}^{\gamma }\right]  \notag \\
&&+\gamma \int d\gamma \int d^{3}r\frac{\partial }{\partial \gamma }\rho
_{N-1}^{\gamma }\left( \mathbf{r}\right) \left( v_{c}\left( \left[ \rho
_{N-1}^{\gamma }\right] ;\mathbf{r}\right) -v_{c}\left( \left[ \rho _{N}%
\right] ;\mathbf{r}\right) \right)  \notag \\
&=&\int d^{3}r^{\prime }\left( \rho _{N}\left( \mathbf{r}^{\prime }\right)
-\rho _{N-1}^{\gamma }\left( \mathbf{r}^{\prime }\right) \right) v_{c}\left( %
\left[ \rho _{N}\right] ;\mathbf{r}\right) .  \label{r1}
\end{eqnarray}

\section{Discussion and summary}

The identities (\ref{id4}) and (\ref{p9}) \cite{Joubert2011b} were derived for $w$%
-representable densities \cite{ParrYang:bk89,DreizlerGross:bk90} and the
proof of Eq. (\ref{apf}) also makes use of the eigenfunctions of a
Hamiltonian. Therefore Eq. (\ref{ID1}) is valid for densities derived from
groundstate wavefunctions of a many-particle Hamiltonian. The assumption was
made that all functional derivatives are well behaved and this implies that
the functionals are defined for non-integer particle numbers \cite{PPLB:82}.
Equation ( (\ref{ID1}) is interesting in that it provides an expression
entirely in terms of correlation functionals and functional derivatives. It
is valid for exact densities and any test of an approximate correlation
functional will only be reliable if exact densities are used. The $\gamma $%
-dependent correlation energy can be derived for any approximate correlation
functional since\cite{Levy1991}%
\begin{equation}
E_{c}^{\gamma }\left[ \rho \right] =\gamma ^{2}E_{c}^{\gamma }\left[ \rho _{%
\frac{1}{\gamma }}\right] ,
\end{equation}%
where $\rho _{\lambda }\left( \mathbf{r}\right) =\lambda ^{3}\rho \left(
\mathbf{\lambda r}\right) $ is the uniformly scaled density. However, to
implement (\ref{ID1}) accurate densities for the $\left( N-1\right) $%
-electron system is needed for all electron interaction strengths between
zero and $\gamma ,$ and this may be difficult to implement in practice.

In summary, a new identity for the correlation functional was derived that
relates the functional and functional derivatives of different densities of
the same Hamiltonian but where the particle number differs by one.

\appendix

\section{}

From Eq. (\ref{ec1})
\begin{eqnarray}
&&\frac{E_{c}^{\gamma }\left[ \rho _{N-1}^{\gamma }\right] -T_{c}^{\gamma }%
\left[ \rho _{N-1}^{\gamma }\right] }{\gamma }  \notag \\
&=&\left\langle \Psi _{\rho _{N-1}^{\gamma }}^{\gamma }\left\vert \hat{V}%
_{ee}\right\vert \Psi _{\rho _{N-1}^{\gamma }}^{\gamma }\right\rangle
-\left\langle \Psi _{\rho _{N-1}^{\gamma }}^{0}\left\vert \hat{V}%
_{ee}\right\vert \Psi _{\rho _{N-1}^{\gamma }}^{0}\right\rangle .
\label{tc1}
\end{eqnarray}%
The derivative of $E_{c}^{\gamma }\left[ \rho _{N-1}^{\gamma }\right] $ with
respect to $\gamma ,$ from definition (\ref{ec1}), can therefore be
expressed as
\begin{eqnarray}
&&\frac{\partial }{\partial \gamma }E_{c}^{\gamma }\left[ \rho
_{N-1}^{\gamma }\right]   \notag \\
&=&\frac{E_{c}^{\gamma }\left[ \rho _{N-1}^{\gamma }\right] -T_{c}^{\gamma }%
\left[ \rho _{N-1}^{\gamma }\right] }{\gamma }  \notag \\
&&+\left\langle \frac{\partial }{\partial \gamma }\Psi _{\rho _{N-1}^{\gamma
}}^{\gamma }\left\vert \hat{T}+\gamma \hat{V}_{ee}\right\vert \Psi _{\rho
_{N-1}^{\gamma }}^{\gamma }\right\rangle   \notag \\
&&+\left\langle \Psi _{\rho _{N-1}^{\gamma }}^{\gamma }\left\vert \hat{T}%
+\gamma \hat{V}_{ee}\right\vert \frac{\partial }{\partial \gamma }\Psi
_{\rho _{N-1}^{\gamma }}^{\gamma }\right\rangle   \notag \\
&&-\left\langle \frac{\partial }{\partial \gamma }\Psi _{\rho _{N-1}^{\gamma
}}^{0}\left\vert \hat{T}+\gamma \hat{V}_{ee}\right\vert \Psi _{\rho
_{N-1}^{\gamma }}^{0}\right\rangle   \notag \\
&&-\left\langle \Psi _{\rho _{N-1}^{\gamma }}^{0}\left\vert \hat{T}+\gamma
\hat{V}_{ee}\right\vert \frac{\partial }{\partial \gamma }\Psi _{\rho
_{N-1}^{\gamma }}^{0}\right\rangle .  \label{tc2}
\end{eqnarray}%
Upon adding and subtracting (c.c. stands for the complex conjugate of the
previous term)
\begin{eqnarray}
&&\left( \left\langle \frac{\partial }{\partial \gamma }\Psi _{\rho
_{N-1}^{\gamma }}^{\gamma }\left\vert \hat{v}_{N-1,\text{ext}}^{\gamma }%
\left[ \rho _{N}\right] \right\vert \Psi _{\rho _{N-1}^{\gamma }}^{\gamma
}\right\rangle +\text{c.c}\right)   \notag \\
&&+\left( \left\langle \left. \frac{\partial }{\partial \gamma }\Psi _{\rho
_{N-1}^{\gamma }}^{\gamma }\right\vert _{\gamma =0}\left\vert \hat{v}_{N-1,%
\text{ext}}^{0}\left[ \rho _{N}\right] \right\vert \Psi _{\rho
_{N-1}^{0}}^{0}\right\rangle +\text{c.c}\right)   \label{tc4}
\end{eqnarray}%
and utilizing the normalization of the wavefunctions which implies that%
\begin{equation}
\frac{\partial }{\partial \gamma }\left\langle \Psi _{\rho _{N-1}^{\gamma
}}^{\gamma }|\Psi _{\rho _{N-1}^{\gamma }}^{\gamma }\right\rangle =0,
\label{ec4}
\end{equation}%
Eq. (\ref{ec2}) becomes%
\begin{eqnarray}
&&\frac{\partial }{\partial \gamma }E_{c}^{\gamma }\left[ \rho
_{N-1}^{\gamma }\right]   \notag \\
&=&\frac{E_{c}^{\gamma }\left[ \rho _{N-1}^{\gamma }\right] -T_{c}^{\gamma }%
\left[ \rho _{N-1}^{\gamma }\right] }{\gamma }  \notag \\
&&-\left\langle \frac{\partial }{\partial \gamma }\Psi _{\rho _{N-1}^{\gamma
}}^{\gamma }\left\vert \hat{v}_{\text{ext}}^{\gamma }\left[ \rho _{N}\right]
\right\vert \Psi _{\rho _{N-1}^{\gamma }}^{\gamma }\right\rangle
-\left\langle \Psi _{\rho _{N-1}^{\gamma }}^{\gamma }\left\vert \hat{v}_{%
\text{ext}}^{\gamma }\left[ \rho _{N}\right] \right\vert \frac{\partial }{%
\partial \gamma }\Psi _{\rho _{N-1}^{\gamma }}^{\gamma }\right\rangle
\notag \\
&&+\left\langle \frac{\partial }{\partial \gamma }\Psi _{\rho _{N-1}^{\gamma
}}^{0}\left\vert \hat{v}_{\text{ext}}^{0}\left[ \rho _{N}\right] \right\vert
\Psi _{\rho _{N-1}^{\gamma }}^{0}\right\rangle +\left\langle \Psi _{\rho
_{N-1}^{\gamma }}^{0}\left\vert \hat{v}_{\text{ext}}^{0}\left[ \rho _{N}%
\right] \right\vert \frac{\partial }{\partial \gamma }\Psi _{\rho
_{N-1}^{\gamma }}^{0}\right\rangle   \notag \\
&&-\gamma \frac{\partial }{\partial \gamma }\left\langle \Psi _{\rho
_{N-1}^{\gamma }}^{0}\left\vert \hat{V}_{ee}\right\vert \Psi _{\rho
_{N-1}^{\gamma }}^{0}\right\rangle .  \label{ec5}
\end{eqnarray}%
This equation can be simplified since
\begin{eqnarray}
&&\left\langle \frac{\partial }{\partial \gamma }\Psi _{\rho _{N-1}^{\gamma
}}^{\gamma }\left\vert \hat{v}_{\text{ext}}^{\gamma }\left[ \rho _{N}\right]
\right\vert \Psi _{\rho _{N-1}^{\gamma }}^{\gamma }\right\rangle
+\left\langle \Psi _{\rho _{N-1}^{\gamma }}^{\gamma }\left\vert \hat{v}_{%
\text{ext}}^{\gamma }\left[ \rho _{N}\right] \right\vert \frac{\partial }{%
\partial \gamma }\Psi _{\rho _{N-1}^{\gamma }}^{\gamma }\right\rangle
\notag \\
&=&\int d^{3}r\frac{\partial \rho _{N-1}^{\gamma }\left( \mathbf{r}\right) }{%
\partial \gamma }\hat{v}_{\text{ext}}^{\gamma }\left( \left[ \rho _{N}\right]
;\mathbf{r}\right) .  \label{ap1}
\end{eqnarray}%
From definition of the exchange energy \cite%
{ParrYang:bk89,DreizlerGross:bk90}
\begin{eqnarray}
\left\langle \Psi _{\rho _{N-1}^{\gamma }}^{0}\left\vert \hat{V}%
_{ee}\right\vert \Psi _{\rho _{N-1}^{\gamma }}^{0}\right\rangle  &=&E_{x}%
\left[ \rho _{N-1}^{\gamma }\right] +U\left[ \rho _{N-1}^{\gamma }\right]
\notag \\
&=&E_{hx}\left[ \rho _{N-1}^{\gamma }\right] ,  \label{ux1}
\end{eqnarray}%
is the sum of the exchange $E_{x}\left[ \rho _{N-1}^{\gamma }\right] $ and
Hartree interaction energy $U\left[ \rho _{N-1}^{\gamma }\right] $ of the $%
\left( N-1\right) $-electron system. The charge density $\rho _{N-1}^{\gamma
}$ is a function of $\gamma $ \cite{ParrYang:bk89}, therefore%
\begin{equation}
\frac{\partial }{\partial \gamma }\left\langle \Psi _{\rho _{N-1}^{\gamma
}}^{0}\left\vert \hat{V}_{ee}\right\vert \Psi _{\rho _{N-1}^{\gamma
}}^{0}\right\rangle =\int d^{3}r\frac{\partial \rho _{N-1}^{\gamma }\left(
\mathbf{r}\right) }{\partial \gamma }v_{hx}\left( \left[ \rho _{N-1}^{\gamma
}\right] ;\mathbf{r}\right) ,  \label{ux2}
\end{equation}%
where
\begin{equation}
v_{hx}\left( \left[ \rho _{N-1}^{\gamma }\right] ;\mathbf{r}\right) =\frac{%
\delta }{\delta \rho _{N-1}^{\gamma }\left( \mathbf{r}\right) }\left( E_{x}%
\left[ \rho _{N-1}^{\gamma }\right] +U\left[ \rho _{N-1}^{\gamma }\right]
\right)   \label{ux3}
\end{equation}%
is the sum of the exchange and Hartree potentials for the $\left( N-1\right)
$-electron system. Using Eqs. (\ref{ap1}), (\ref{ux2}), (\ref{e1}) and the
fact that $\left\vert \Psi _{\rho _{N-1}^{\gamma }}^{\gamma }\right\rangle $
and $\left\vert \Psi _{\rho _{N-1}^{\gamma }}^{0}\right\rangle $ yield the
same density $\rho _{N-1}^{\gamma }$, Eq. (\ref{ec5}) can be expressed as%
\begin{eqnarray}
&&\frac{\partial }{\partial \gamma }E_{c}^{\gamma }\left[ \rho
_{N-1}^{\gamma }\right]   \notag \\
&=&\frac{E_{c}^{\gamma }\left[ \rho _{N-1}^{\gamma }\right] -T_{c}^{\gamma }%
\left[ \rho _{N-1}^{\gamma }\right] }{\gamma }  \notag \\
&&+\int d^{3}r\frac{\partial \rho _{N-1}^{\gamma }\left( \mathbf{r}\right) }{%
\partial \gamma }\left( v_{c}^{\gamma }\left( \left[ \rho _{N}\right] ;%
\mathbf{r}\right) +\gamma v_{hx}\left( \left[ \rho _{N}\right] ;\mathbf{r}%
\right) -\gamma v_{hx}\left( \left[ \rho _{N-1}^{\gamma }\right] ;\mathbf{r}%
\right) \right)   \label{apf}
\end{eqnarray}


\begin{thebibliography}{19}
\expandafter\ifx\csname natexlab\endcsname\relax\def\natexlab#1{#1}\fi
\expandafter\ifx\csname bibnamefont\endcsname\relax
  \def\bibnamefont#1{#1}\fi
\expandafter\ifx\csname bibfnamefont\endcsname\relax
  \def\bibfnamefont#1{#1}\fi
\expandafter\ifx\csname citenamefont\endcsname\relax
  \def\citenamefont#1{#1}\fi
\expandafter\ifx\csname url\endcsname\relax
  \def\url#1{\texttt{#1}}\fi
\expandafter\ifx\csname urlprefix\endcsname\relax\def\urlprefix{URL }\fi
\providecommand{\bibinfo}[2]{#2}
\providecommand{\eprint}[2][]{\url{#2}}

\bibitem[{\citenamefont{Kohn and Sham}(1965)}]{KohnSham:65}
\bibinfo{author}{\bibfnamefont{W.}~\bibnamefont{Kohn}} \bibnamefont{and}
  \bibinfo{author}{\bibfnamefont{L.~J.} \bibnamefont{Sham}},
  \bibinfo{journal}{Phys. Rev. A} \textbf{\bibinfo{volume}{140}},
  \bibinfo{pages}{1133} (\bibinfo{year}{1965}).

\bibitem[{\citenamefont{Hohenberg and Kohn}(1964)}]{HohenbergKohn:64}
\bibinfo{author}{\bibfnamefont{P.}~\bibnamefont{Hohenberg}} \bibnamefont{and}
  \bibinfo{author}{\bibfnamefont{W.}~\bibnamefont{Kohn}},
  \bibinfo{journal}{Phys. Rev. B} \textbf{\bibinfo{volume}{136}},
  \bibinfo{pages}{864} (\bibinfo{year}{1964}).
  
  \bibitem[{\citenamefont{Parr and Yang}(1989)}]{ParrYang:bk89}
\bibinfo{author}{\bibfnamefont{R.~G.} \bibnamefont{Parr}} \bibnamefont{and}
  \bibinfo{author}{\bibfnamefont{W.}~\bibnamefont{Yang}},
  \emph{\bibinfo{title}{Density Functional Theory of Atoms and Molecules}}
  (\bibinfo{publisher}{Oxford University Press}, \bibinfo{address}{New York},
  \bibinfo{year}{1989}).

\bibitem[{\citenamefont{Dreizler and Gross}(1990)}]{DreizlerGross:bk90}
\bibinfo{author}{\bibfnamefont{R.~M.} \bibnamefont{Dreizler}} \bibnamefont{and}
  \bibinfo{author}{\bibfnamefont{E.~K.~U.} \bibnamefont{Gross}},
  \emph{\bibinfo{title}{Density Functional Theory}}
  (\bibinfo{publisher}{Springer-Verlag}, \bibinfo{address}{Berlin},
  \bibinfo{year}{1990}).

\bibitem[{\citenamefont{Ruzsinszky and Perdew}(2011)}]{AdriennRuzsinszky2011}
\bibinfo{author}{\bibfnamefont{A.}~\bibnamefont{Ruzsinszky}} \bibnamefont{and}
  \bibinfo{author}{\bibfnamefont{J.~P.} \bibnamefont{Perdew}},
  \bibinfo{journal}{Computational and Theoretical Chemistry}
  \textbf{\bibinfo{volume}{963}}, \bibinfo{pages}{2} (\bibinfo{year}{2011}).

\bibitem[{\citenamefont{Perdew et~al.}(2010)\citenamefont{Perdew, Ruzsinszky,
  Csonka, Constantin, and Sun}}]{JohnP.Perdew2009}
\bibinfo{author}{\bibfnamefont{J.~P.} \bibnamefont{Perdew}},
  \bibinfo{author}{\bibfnamefont{A.}~\bibnamefont{Ruzsinszky}},
  \bibinfo{author}{\bibfnamefont{G.~I.} \bibnamefont{Csonka}},
  \bibinfo{author}{\bibfnamefont{L.~A.} \bibnamefont{Constantin}},
  \bibnamefont{and} \bibinfo{author}{\bibfnamefont{J.}~\bibnamefont{Sun}},
  \bibinfo{journal}{Phys. Rev. Lett.} \textbf{\bibinfo{volume}{103}},
  \bibinfo{pages}{026403} (\bibinfo{year}{2010}).

\bibitem[{\citenamefont{Csonka et~al.}(2009)\citenamefont{Csonka, Perdew,
  Ruzsinszky, Philipsen, Leb\`egue, Paier, Vydrov, and
  \'Angy\'an}}]{GaborI.Csonka2009a}
\bibinfo{author}{\bibfnamefont{G.~I.} \bibnamefont{Csonka}},
  \bibinfo{author}{\bibfnamefont{J.~P.} \bibnamefont{Perdew}},
  \bibinfo{author}{\bibfnamefont{A.}~\bibnamefont{Ruzsinszky}},
  \bibinfo{author}{\bibfnamefont{P.~H.~T.} \bibnamefont{Philipsen}},
  \bibinfo{author}{\bibfnamefont{S.}~\bibnamefont{Leb\`egue}},
  \bibinfo{author}{\bibfnamefont{J.}~\bibnamefont{Paier}},
  \bibinfo{author}{\bibfnamefont{O.~A.} \bibnamefont{Vydrov}},
  \bibnamefont{and} \bibinfo{author}{\bibfnamefont{J.~G.}
  \bibnamefont{\'Angy\'an}}, \bibinfo{journal}{Phys. Rev. B}
  \textbf{\bibinfo{volume}{79}}, \bibinfo{pages}{155107}
  (\bibinfo{year}{2009}).

\bibitem[{\citenamefont{Staroverov et~al.}(2004)\citenamefont{Staroverov,
  Scuseria, Tao, and Perdew}}]{Staroverov2004}
\bibinfo{author}{\bibfnamefont{V.~N.} \bibnamefont{Staroverov}},
  \bibinfo{author}{\bibfnamefont{G.~E.} \bibnamefont{Scuseria}},
  \bibinfo{author}{\bibfnamefont{J.}~\bibnamefont{Tao}}, \bibnamefont{and}
  \bibinfo{author}{\bibfnamefont{J.~P.} \bibnamefont{Perdew}},
  \bibinfo{journal}{Phys. Rev. B} \textbf{\bibinfo{volume}{69}},
  \bibinfo{pages}{075102} (\bibinfo{year}{2004}).

\bibitem[{\citenamefont{Tao et~al.}(2003)\citenamefont{Tao, Perdew, Staroverov,
  and Scuseria}}]{TaoPerdew03}
\bibinfo{author}{\bibfnamefont{J.}~\bibnamefont{Tao}},
  \bibinfo{author}{\bibfnamefont{J.~P.} \bibnamefont{Perdew}},
  \bibinfo{author}{\bibfnamefont{V.~N.} \bibnamefont{Staroverov}},
  \bibnamefont{and} \bibinfo{author}{\bibfnamefont{G.~E.}
  \bibnamefont{Scuseria}}, \bibinfo{journal}{Phys. Rev. Lett.}
  \textbf{\bibinfo{volume}{91}}, \bibinfo{pages}{146401}
  (\bibinfo{year}{2003}).

\bibitem[{\citenamefont{Perdew et~al.}(2005)\citenamefont{Perdew, Ruzsinszky,
  and Tao}}]{JohnP.Perdew2005}
\bibinfo{author}{\bibfnamefont{J.~P.} \bibnamefont{Perdew}},
  \bibinfo{author}{\bibfnamefont{A.}~\bibnamefont{Ruzsinszky}},
  \bibnamefont{and} \bibinfo{author}{\bibfnamefont{J.}~\bibnamefont{Tao}},
  \bibinfo{journal}{J. Chem. Phys.} \textbf{\bibinfo{volume}{123}},
  \bibinfo{pages}{062201} (\bibinfo{year}{2005}).

\bibitem[{\citenamefont{Harris and Jones}(1974)}]{HarrisJones:74}
\bibinfo{author}{\bibfnamefont{J.}~\bibnamefont{Harris}} \bibnamefont{and}
  \bibinfo{author}{\bibfnamefont{R.~O.} \bibnamefont{Jones}},
  \bibinfo{journal}{J. Phys. F} \textbf{\bibinfo{volume}{4}},
  \bibinfo{pages}{1174} (\bibinfo{year}{1974}).

\bibitem[{\citenamefont{Langreth and Perdew}(1975)}]{LangrethPerdew:75}
\bibinfo{author}{\bibfnamefont{D.~C.} \bibnamefont{Langreth}} \bibnamefont{and}
  \bibinfo{author}{\bibfnamefont{J.~P.} \bibnamefont{Perdew}},
  \bibinfo{journal}{Solid State Comm.} \textbf{\bibinfo{volume}{17}},
  \bibinfo{pages}{1425} (\bibinfo{year}{1975}).

\bibitem[{\citenamefont{Langreth and Perdew}(1977)}]{LangrethPerdew:77}
\bibinfo{author}{\bibfnamefont{D.~C.} \bibnamefont{Langreth}} \bibnamefont{and}
  \bibinfo{author}{\bibfnamefont{J.~P.} \bibnamefont{Perdew}},
  \bibinfo{journal}{Phys. Rev. B} \textbf{\bibinfo{volume}{15}},
  \bibinfo{pages}{2884} (\bibinfo{year}{1977}).

\bibitem[{\citenamefont{Gunnarson and Lundqvist}(1976)}]{GunnarsonLundqvist:76}
\bibinfo{author}{\bibfnamefont{O.}~\bibnamefont{Gunnarson}} \bibnamefont{and}
  \bibinfo{author}{\bibfnamefont{B.~I.} \bibnamefont{Lundqvist}},
  \bibinfo{journal}{Phys. Rev. B} \textbf{\bibinfo{volume}{13}},
  \bibinfo{pages}{4274} (\bibinfo{year}{1976}).

\bibitem[{\citenamefont{Levy}(1979)}]{Levy:79}
\bibinfo{author}{\bibfnamefont{M.}~\bibnamefont{Levy}}, \bibinfo{journal}{Natl.
  Acad. Sci. USA} \textbf{\bibinfo{volume}{76}}, \bibinfo{pages}{6062}
  (\bibinfo{year}{1979}).

\bibitem[{\citenamefont{Levy and Perdew}(1985)}]{LevyPerdew:85}
\bibinfo{author}{\bibfnamefont{M.}~\bibnamefont{Levy}} \bibnamefont{and}
  \bibinfo{author}{\bibfnamefont{J.~P.} \bibnamefont{Perdew}},
  \bibinfo{journal}{Phys. Rev. A} \textbf{\bibinfo{volume}{32}},
  \bibinfo{pages}{2010} (\bibinfo{year}{1985}).

\bibitem[{\citenamefont{G{\"o}rling and Levy}(1993)}]{GorlingLevy:93}
\bibinfo{author}{\bibfnamefont{A.}~\bibnamefont{G{\"o}rling}} \bibnamefont{and}
  \bibinfo{author}{\bibfnamefont{M.}~\bibnamefont{Levy}},
  \bibinfo{journal}{Phys. Rev. B} \textbf{\bibinfo{volume}{47}},
  \bibinfo{pages}{13105} (\bibinfo{year}{1993}).



\bibitem[{\citenamefont{Perdew et~al.}(1982)\citenamefont{Perdew, Parr, Levy,
  and Balduz}}]{PPLB:82}
\bibinfo{author}{\bibfnamefont{J.}~\bibnamefont{Perdew}},
  \bibinfo{author}{\bibfnamefont{R.}~\bibnamefont{Parr}},
  \bibinfo{author}{\bibfnamefont{M.}~\bibnamefont{Levy}}, \bibnamefont{and}
  \bibinfo{author}{\bibfnamefont{J.}~\bibnamefont{Balduz}},
  \bibinfo{journal}{Phys. Rev. Lett.} \textbf{\bibinfo{volume}{49}},
  \bibinfo{pages}{1691} (\bibinfo{year}{1982}).

\bibitem[{\citenamefont{Joubert}(2011)}]{Joubert2011b}
\bibinfo{author}{\bibfnamefont{D.~P.} \bibnamefont{Joubert}},
  \bibinfo{journal}{arXiv:1108.1094v2 [cond-mat.mtrl-sci]}
  (\bibinfo{year}{2011}).
  
\bibitem[{\citenamefont{Levy}(1991)}]{Levy1991}
\bibinfo{author}{\bibfnamefont{M.}~\bibnamefont{Levy}}, \bibinfo{journal}{Phys.
  Rev. A} \textbf{\bibinfo{volume}{43}}, \bibinfo{pages}{4637}
  (\bibinfo{year}{1991}).  

\end{thebibliography}
\end{document}